\documentstyle[preprint,prl,aps]{revtex}

\newcommand{\tj}{$t$-$J$}
\newcommand{\dxy}{$d_{x^2-y^2}$}
\newcommand{\etal}{{\it et al}}
\newcommand{\ij}{\langle ij\rangle}
\renewcommand{\S}{{\vec S}}  

\begin{document}
\draft
\title{\dxy\  Pair Domain Walls}
\author{ Steven R.\ White$^1$ and D.J.\ Scalapino$^2$}
\address{ 
$^1$Department of Physics and Astronomy,
University of California,
Irvine, CA 92697
}
\address{ 
$^2$Department of Physics,
University of California,
Santa Barbara, CA 93106
}
\date{\today}
\maketitle
\begin{abstract}

Using the density matrix renormalization group,
we study domain wall structures in the \tj\ model at a
hole doping of $x={1\over8}$.
We find that the domain walls are composed of \dxy\  pairs and that the
regions between the domain walls have antiferromagnetic correlations that
are $\pi$ phase shifted across a domain wall.
At $x={1\over8}$, the hole filling corresponds to one hole per two
domain wall unit cells.
When the pairs in a domain wall are pinned by an external field, the 
\dxy\ pairing response is suppressed, but when the pinning is weakened,
\dxy\ pair-field correlations can develop.

\end{abstract}

\narrowtext

In the low temperature tetragonal (LTT) phase of 
La$_{1.6-x}$Nd$_{0.4}$Sr$_x$CuO$_4$, the tilt pattern of the
CuO$_6$ octahedra form lines of displaced oxygens parallel to the
Cu-O bond directions. These lines are rotated by 90$^\circ$ between
adjacent layers. At a filling of $x={1\over8}$, superconductivity is
suppressed and neutron scattering studies\cite{TN,T} reveal a striped domain
wall ordering of holes and spins which is believed to be
commensurately locked by the tilt distortion of the lattice. One
model for this striped order\cite{TN,T} is illustrated in Fig.~1(a).
Here the charge domain walls are shown running vertically and
centered along the Cu-O-Cu legs, although the phase information required
to determine whether the domains should be leg centered or bond centered
(centered between two legs) is not known. As shown, the domains are
separated by four Cu-O-Cu spacings and for $x={1\over8}$ contain
one hole per two $4\times 1$ domain wall unit cells. 
This latter feature is at odds with one-electron Hartree-Fock
calculations\cite{Z,P,S,V,IL,ZO} which predict a domain wall filling of one
hole per domain wall unit cell. The spins in the regions between the
walls are antiferromagnetically correlated with a $\pi$ phase shift
across a domain wall.
When $x\neq{1\over8}$, superconductivity is found to
coexist with a weakened domain wall ordering, suggesting a close
connection between the two.

Here we present results of numerical density matrix renormalization
group calculations for a \tj\ model with a hole doping $x={1\over8}$.
We find evidence for domain walls with $\pi$ phase-shifted
antiferromagnetic regions separating the walls, and with a filling
of one hole per two $4\times 1$ domain wall unit cells.
In contrast to Fig. 1(a), however, the domain walls are bond
centered, and are made up of \dxy\  pairs of holes.
We find that just as for the two-leg and four-leg ladder 
problems,\cite{N,fourchain} there is a competition between a charge
density wave (CDW) phase and a superconducting pair phase. 
The Coulomb interaction, arising from
the 90$^\circ$ rotated domain walls on adjacent planes in the LTT
phase\cite{T,K}, acts to pin the pairs within a domain 
wall for $x={1\over8}$.
We show that when this pinning is weakened, \dxy\  pair field
correlations develop.

The \tj\ Hamiltonian we will study is given by
\begin{equation}
H= -\sum_{\ij s} t\, P_G 
\left(c^\dagger_{i,s}c_{j,s} + c^\dagger_{j,s}c_{i,s} \right) P_G
+ \sum_{\ij} J\, \left(\S_i\cdot\S_j - {1\over4}n_in_j \right).
\end{equation}
Here $\ij$ are near-neighbor sites, $s$ is a spin index, $\S_i =
c^\dagger_{i,s} {\sigma}_{s,s'}c_{i,s'}$ and $n_i=
c^\dagger_{i\uparrow}c_{i\uparrow} +
c^\dagger_{i\downarrow}c_{i\downarrow}$, with $c^\dagger_{is}$
($c_{is}$) an operator which creates (destroys) an electron at site $i$
with spin $s$.
The near-neighbor hopping interaction is $t$, the near-neighbor
exchange interaction is $J$, and
the Gutzwiller projection operator $P_G$ excludes configurations with
doubly occupied sites. We refer to the Cu-Cu lattice spacing as $a$
and measure energies in units of $t$.

Our calculations were carried out using a density matrix renormalization 
group\cite{dmrg} (DMRG) algorithm in which transformation
matrices were stored and used to construct the initial state for
each superblock diagonalization\cite{cavo}.  
We present results here for $8\times8$ and $4\times18$ doped systems.
For the more difficult $8\times8$ system,
typically 10 equilibration sweeps were made through the lattice, keeping of
order $10^3$ states per block on the final sweep. 
The transformation matrices were used
to calculate the ground state expectation values of the desired
operators at the end of the calculation. 
DMRG is extremely accurate for narrow systems, but its accuracy
decreases rapidly as the width increases, and  consequently, our
$8\times8$ systems may represent the most ambitious DMRG 
calculations to date. Truncation errors were approximately
$0.0003$, which we consider to be at the limits of acceptability. In
addition, the calculations sometimes got stuck in metastable hole
configurations, in cases where two or more states had a very close
separation in energy but corresponded to completely different hole
configurations. In these cases it was necessary to perform several
calculations, with the holes initially localized in different
configurations, using the total energy to choose between different
runs. (In contrast, the spin configuration equilibrated more
readily.) Nevertheless, the domain wall configurations shown here
were robust as long as the initial hole configurations 
were reasonably close to the final configurations displayed.
In particular, three of the most robust features emerging from a
variety of runs were that the domain walls are bond centered,
composed of \dxy\  pairs, with $\pi$ phase shifted antiferromagnetic
regions separating the walls.

Figure 1(b) shows the charge and spin density in the ground state of the
$8\times8$ system with $J/t=0.35$ and 8 holes, corresponding to a
filling $x={1\over8}$. Periodic boundary conditions were used
in the $y$-direction, and open boundary conditions in the
$x$-direction. Along the left and right edges of the system a 
small staggered magnetic field of $0.05t$ was applied. The boundary
conditions and the edge staggered field serve to orient and pin the domain
walls in the configuration shown. In an LTT phase, the increased Cu-O 
bond length in one direction would act to orient the domain walls
through an anisotropic hopping and exchange, $t_y=\gamma t_x$ and 
$J_y = \gamma^2 J_x$, with $\gamma \approx 1$, but for simplicity 
we have used $\gamma=1$ throughout. The staggered edge field further
acts to pick a direction for the spin order, which allows direct
measurement of the spin configurations and reduces truncation errors
in the DMRG calculation.
The charge density in the $x$-direction shows a strong modulation
with period 4$a$, and the filling of the wall is consistent with
that shown in Fig. 1(a).
In addition, a small charge density modulation, also with period of
4$a$, is present in the $y$ direction\cite{Z}. This modulation
is pinned in the $y$ direction despite periodic boundary conditions
by the truncation errors in the calculation. We cannot determine
whether an exact calculation of the two-dimensional $t$-$J$ lattice
would have small pinned charge density modulations along the walls, but it
seems clear that the system is near a CDW instability, with period 4$a$,
along the walls.
The spin response $\langle S^z_i\rangle$ is also shown in
Fig. 1(b) and corresponds to antiferromagnetic correlations which have a
$\pi$ phase shift across the bond domain wall. This $\pi$ phase
shift was purely the result of the simulation, not enhanced in any way
by boundary or initial conditions.

In Fig.~1(c), the black circles show the most probable configuration of
the eight holes obtained by maximizing the ground state expectation
value of
\begin{equation}
P(\ell_1,\ell_2,\ldots,\ell_8) = \prod_i p(\ell_i),
\end{equation}
with $p(\ell) = (1-n_{\ell\uparrow})(1-n_{\ell\downarrow})$ the hole
projection operator for the $\ell^{\rm th}$ lattice site.
The thickness of the lines connecting various sites denotes the strength
of the exchange field $\langle\S_i\cdot\S_j\rangle$ when the holes
occupy the most probable configuration, consisting of four pairs.
Note that the most probable configuration of a pair is a diagonal
configuration, with a strong exchange bond running diagonally 
between the spins adjacent to the holes. This configuration is
the most likely configuration of a pair in a variety of \tj\
clusters\cite{holestructures,fourchain}.

From a strong coupling point of view, the pairing can be viewed as
arising from a compromise in which two holes locally arrange
themselves so as to minimize the disturbance of the background
exchange energy while at the same time lowering their kinetic
energy.  In the unphysical regime of $J>t$, the
holes would tend to be near neighbors to reduce the
number of broken exchange bonds.  In the physical region $J<t$, the
kinetic energy plays an increasingly important role so that, as
shown in Fig.~1(c), the most probable pair configuration has the
holes sitting on diagonal sites with strong singlet
correlations on the other diagonal. In this case, four of the eight
one-electron hops leads to a configuration with a near-neighbor
exchange bond which stabilizes the pair.
As previously discussed,\cite{Trug,holestructures} the pair structure 
is such that it has an overlap with the undoped 
antiferromagnetic background through a hole pair field operator
which has \dxy\ symmetry. Here we will see the \dxy\ structure of
the pairs from the pair field response discussed below.

In the LTT phase, the CuO$_6$ tilt structure causes the domain walls to
be perpendicular in adjacent planes\cite{TN}. This gives rise to
an electrostatic potential, with a period of 4$a$, along the domain wall
\cite{K}.  For $x=1/8$, this corresponds to the CDW instability of the \dxy\ 
domain wall and can lead to a pinning of the pairs along the 
wall.  Here we have modeled this effect by adding a spatially
varying site potential $\Delta V =0.1t$ on the $8\times8$ lattice
shown in Fig. 2.  The sites with the extra potential $\Delta V$  are
shown by the shaded rectangles, which also
indicate the domain walls in the adjacent planes.
The potential pins the pairs, forming a CDW lattice of pairs.

The charge and spin structure factors 
\begin{equation}
S_c({\bf q}) = \frac{1}{64} \sum_\ell e^{i {\bf q} \cdot \vec \ell}
\langle n_{\vec l \uparrow} + n_{\vec l \downarrow} \rangle
\end{equation}
and 
\begin{equation}
S_\sigma({\bf q}) = \frac{1}{64} \sum_\ell e^{i {\bf q} \cdot \vec \ell}
\langle n_{\vec l \uparrow} - n_{\vec l \downarrow} \rangle
\end{equation}
for this lattice are also shown in Fig.~2. The intensity in
$S_c({\bf q})$ at $(\frac{\pi}{2a},0)$ is expected for vertical charged
domain walls. The intensity at $(\frac{\pi}{2a},\frac{\pi}{2a})$ 
and $(0,\frac{\pi}{2a})$
reflect the pair correlations along the domain walls. With the
90$^\circ$ rotations of the domain walls from layer to layer, it is
the $(\frac{\pi}{2a},\frac{\pi}{2a})$ peak that would be important to observe.
The spin structure factor at $(\frac{3\pi}{4a},\frac{\pi}{a})$ reflects 
structure of the $\pi$-phase shifted antiferromagnetic regions. The
amplitude of the third harmonic intensities is much weaker.

At dopings away from $x=1/8$, the 
La$_{1.6-x}$Nd$_{0.4}$Sr$_x$CuO$_4$ system becomes superconducting,
and the size of the tilt modulation as well as the intensity of the
magnetic Bragg peaks decreases\cite{T}.  This suggests that when the pinning
is weakened, either through a reduced pinning potential or through
a mismatch in the periods of the pinning potential and the CDW
instability, stripe order and superconducting pairing can coexist. 
As a test of this, we have used a $4\times18$ cluster to model a
single, longer domain wall, with a variable pinning potential
$\Delta V$ which acts on the sites in the shaded regions shown in Fig.~3(a).
This cluster has open boundary conditions on all sides with a
staggered magnetic field of magnitude 0.05$t$ applied
along the top and bottom edges. The magnetic field has a
$\pi$ phase shift between the edges in order to mimic the single
domain wall structure shown in Fig.~2(a). 
In order to measure the tendency for superconductivity, a weak pair field 
proximity effect term 
\begin{equation}
H_1 = d\sum_i (\Delta^\dagger_{i,i+\hat y} + \Delta_{i,i+\hat y} )
\end{equation}
was added to the Hamiltonian. Here
\begin{equation}
\Delta_{i,i+\hat y}^\dagger  = 
c^\dagger_{i,\uparrow} c^\dagger_{i+\hat y,\downarrow} +
c^\dagger_{i+\hat y,\uparrow} c^\dagger_{i,\downarrow}
\end{equation}
adds a singlet electron pair to sites $i$ and $i+\hat y$. 
Note that the interaction $H_1$ couples equally to \dxy-like and
extended $s$-wave-like pairs.
That is, it does not distinguish whether the internal structure of the
pair field has a change in sign for the singlet components in the
$y$-direction relative to those in the $x$-direction.
In order to include this term, rather than use 
the number of electrons $N$ as a good quantum number to break up the 
Hilbert space, $N$ modulo 2 was used.  (Total $S_z$ was conserved in
the usual fashion.)
We then measured the strength of the pair fields in the ground state
in both the $x$ and $y$ directions,
$\langle \Delta_{i,i+\hat x}\rangle$ and 
$\langle \Delta_{i,i+\hat y}\rangle$ for all sites $i$. 
The charge and spin structure of the $4\times18$ cluster with 8 holes is
shown in Fig.~3(a) for $\Delta\,V=0.05t$.
In Fig.~3(b), the pair field strength is shown by the width of the
lines, and the sign of the field is indicated by
the type of line, dashed or solid. A relative sign difference
between the $x$ and $y$ directions indicates \dxy\ pairing. In
Fig.~3(b) we see a significant \dxy\ pair response coexisting
with a modest charge density wave.
In Fig.~3(c), we show the pair field 
$\langle\Delta_{\rm mid}\rangle$ averaged over
all the $y$-bonds in the middle four rungs as a function of $\Delta V$. 
The suppression of pairing by the CDW is evident. We expect that a
larger, two dimensional array of
domain walls would show a more enhanced response versus $\Delta V$.

These calculations show that holes doped into a \tj\ lattice can form
domain walls of pairs.
For a filling $x={1\over8}$, these walls have an average filling of
one hole per two domain wall unit cells, and there is a tendency for
the pairs to form a pinned CDW structure. If the pinning is
weakened, the pairs fluctuate, developing phase coherence, and \dxy\ 
superconducting correlations appear.
The idea that the domain walls are made up of pairs provides a natural
explanation of the special filling $x={1\over8}$ and the intimate
relationship between the stripes and superconductivity.
The pairs give rise to a bond centered charge density periodicity of 
$4a$ along a wall which should give a Bragg peak at
$(\frac{\pi}{2a},\frac{\pi}{2a})$.  Other evidence of the pair
structure of the domain wall should be seen in NMR measurements.

The structure we have discussed differs from the one-electron mean
field domain walls found in Hartree-Fock\cite{Z,P,S,V,IL,ZO} and
Gutzwiller variational calculations\cite{GL} in both its filling and
its relationship to pairing.  It also differs from the
one-dimensional large $U/t$ Hubbard model of a domain wall recently
discussed by Nayak and Wilczek\cite{NW}.  At $J/t=0.35$, our system
is well away from the phase separation regime, and we have no
long-range intraplanar Coulomb interaction as in the frustrated
phase separation domain wall model of Kivelson and Emery\cite{K}.
The singlet stripe phases discussed by
Tsunetsugu\cite{TTR,note} and the competing CDW-pairing phases in the 2-
and 4-leg ladder systems recently discussed\cite{N,fourchain} have
some of the features found in this present study.  In particular,
they have the interplay of \dxy\ superconducting pairing and 
charge density wave order.  Clearly, it will be important to understand
if this is the behavior that underlies the structure of the domain
walls in La$_{1.6-x}$Nd$_{0.4}$Sr$_x$CuO$_4$.

\section*{Acknowledgements}

We would like to thank J.~Tranquada for discussions of his experimental
results, and J. Lawrence for helpful discussions.  
SRW acknowledges support from the NSF under 
Grant No. DMR-9509945, and DJS acknowledges support from the
Department of Energy under grand DE-FG03-85ER45197, and from the
Program on Correlated Electrons at the Center for Material Science
at Los Alamos National Laboratory.

\newpage

\begin{figure}
\caption{(a) Spin and hole structure suggested in 
Ref. [1] to
account for neutron scattering experiments.
(b) Hole density and spin moments for an $8\times8$ \tj\  model.
The diameter of the gray holes is proportional to the hole density
$1-\langle n_i\rangle$,
and the length of the arrows is proportional to $\langle S^z_i \rangle$,
according to the scales shown.
(c) For the same system, the exchange field strength 
$-\langle\S_i\cdot\S_j\rangle$ is given by the width of the lines 
according to the scale shown, when the holes (black dots) occupy their most
probable configuration. In addition to showing all nearest 
neighbor exchange bonds, we show next nearest neighbor correlations
about each hole if those correlations are antiferromagnetic, 
$\langle\S_i\cdot\S_j\rangle < 0$.
}
\end{figure}
\begin{figure}
\caption{
(a) Hole density and spin moments for an $8\times8$ \tj\  model 
with a static potential $\Delta V=0.05t$ applied to the sites in the
shaded rectangles.
(b) Charge structure factor $S_c({\bf q})$ expected for a 2D system,
obtained by periodically repeating the pattern shown in (a).
(c) Spin structure factor $S_\sigma({\bf q})$. Here we have measured
$\ell$ from the center of a domain wall. In both (b) and (c), the
area of the circle is proportional to the structure factor, 
open circles denote negative values, and the axes are labeled in
units of $1/a$.
}
\end{figure}
\begin{figure}
\caption{
A single domain wall, modeled by a $4\times18$ system with left and 
right edge $\pi$-phase shifted antiferromagnetic fields $h=0.05t$, 
with a static
potential $\Delta V = 0.05t$, and with an applied proximity
affect field $d=0.03t$.
(a) Hole density and spin moments. (b) Pair field strengths on each
nearest neighbor link. (c) The average pairing strength on the
$y$ links for the central 4 rungs, as a function of $\Delta V$.
}
\end{figure}

\begin{references}
\bibitem{TN}J.M.~Tranquada \etal, Nature {\bf 375}, 561 (1995);
\prb {\bf 54}, 7489 (1996).

\bibitem{T}J.M.~Tranquada et. al., preprint (cond-mat/9608048).

\bibitem{Z}J.~Zaanen and O.~Gunnarsson, \prb {\bf 40}, 7391 (1989).

\bibitem{P}D.~Poilblanc and T.M.~Rice, \prb {\bf 39}, 9749 (1989).

\bibitem{S}H.J.~Schulz, J.~Physique, {\bf 50}, 2833 (1989).

\bibitem{V}J.A.~Verg\'es \etal, \prb {\bf 43}, 6099 (1991).

\bibitem{IL}M.~Inui and P.B.~Littlewood, \prb {\bf 44}, 4415 (1991).

\bibitem{ZO}J.~Zaanen and A.M.~Oles, Ann.\ Physik {\bf 5}, 224,
(1996).

\bibitem{N}R.M.~Noack, S.R.~White, and D.J.~Scalapino, \prl {\bf 73}, 882 
(1994).

\bibitem{fourchain}S.R.~White and D.J.~Scalapino, preprint
(cond-mat/9608138). 

\bibitem{K}S.A.~Kivelson and V.J.~Emery, p.~619 in {\it Proc.
``Strongly Correlated Electronic Materials: The Los Alamos Symposium
1993,''} K.S.~Bedell \etal, eds. (Addison Wesley, Redwood City, Ca.,
1994); S.A.~Kivelson and V.J.~Emery, preprint (cond-mat/9603009).

\bibitem{dmrg} S.R. White, \prl {\bf 69}, 2863 (1992),
\prb {\bf 48}, 10345 (1993).

\bibitem{cavo} S.R. White, \prl {\bf 77}, 3633 (1996).

\bibitem{holestructures} S. R. White and D.J. Scalapino,
preprint (cond-mat/9605143).

\bibitem{Trug}D.J.~Scalapino and S.~Trugman,  to appear in {\sl 
Phil.\ Mag.\ B} (1996).

\bibitem{GL}T.~Giamarchi and C.~Lhuillier, \prb {\bf 42}, 10641 (1990).

\bibitem{NW}C.~Nayak and F.~Wilczek, preprint (cond-mat/9602112).

\bibitem{TTR} H. Tsunetsugu, M. Troyer, and T.M. Rice,
\prb {\bf 51}, 16456 (1995).

\bibitem{note} The region between the walls in our system appears
to be a low energy two-leg mostly-undoped ladder of the typed discussed
previously\cite{TTR,holestructures}. However, in the system we have 
discussed here, there is a
significant $\pi$ phase shifted exchange coupling between ladders,
mediated by the domain walls. This
exchange coupling reduces, and may eliminate, the spin gap.

\end{references}
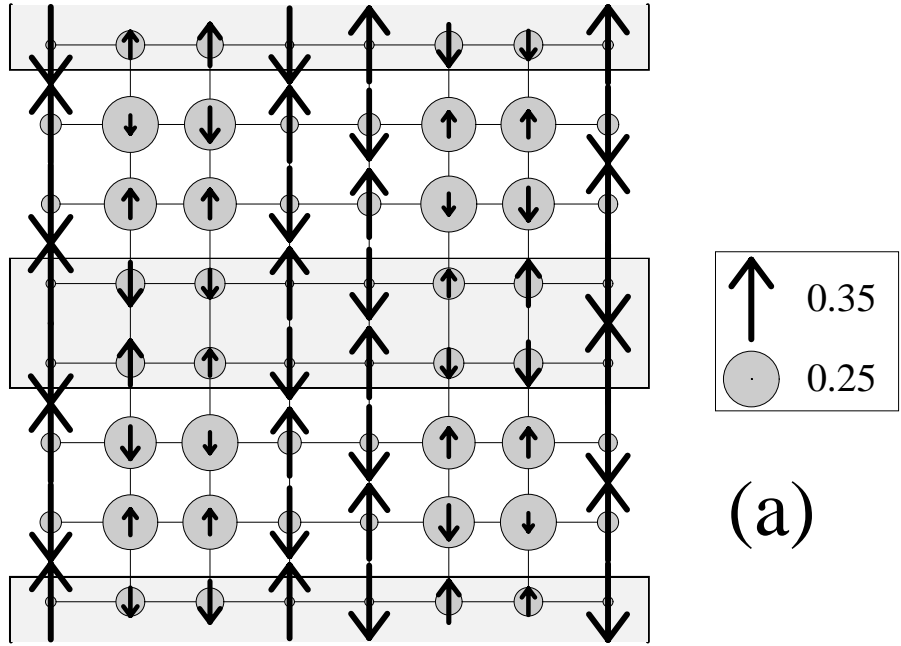
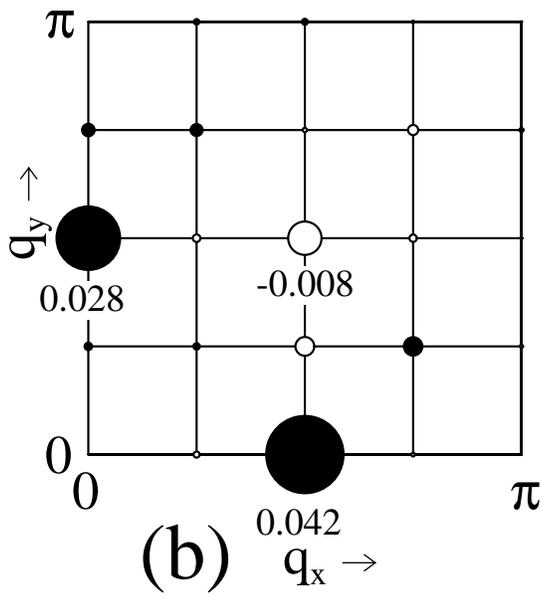
\end{document}